\documentstyle[%
twocolumn,%
aps,%
prl,%
psfig%
]{revtex}

\begin{document}


\draft

\title{Suppression of the Coulomb Interaction Contribution to the 
Conductance\\ 
by a Parallel Magnetic Field}

\author{S.~G.~den~Hartog, S.~J.~van~der~Molen, B.~J.~van~Wees, 
and T.~M.~Klapwijk}

\address{Department of Applied Physics and Material Science Center,
University of Groningen,\\ Nijenborgh 4, 9747 AG Groningen,
The Netherlands.}

\author{G. Borghs}

\address{Interuniversity Micro Electronics Centre, Kapeldreef 75,
B-3030 Leuven, Belgium}

\date{\today}

\maketitle

\begin{abstract}
The Coulomb interaction contribution to the conductance 
is investigated in a phase-coherent disordered 
2-dimensional electron gas, 
which resistance can be varied by an overall gate electrode.
Its magnitude of $\delta G_{EEI} \simeq -0.3 e^2/h$ is obtained by 
applying a bias voltage to suppress the Coulomb anomaly.
In contrast to theoretical predictions, $\delta G_{EEI}$
is suppressed by a parallel magnetic field. 
The zero-bias magnetoresistance exhibits reproducible fluctuations
in perpendicular magnetic fields on a field scale much larger than that 
expected for universal conductance fluctuations,
which might be attributed to fluctuations in the Coulomb interaction 
contribution.
\end{abstract}



Most of the transport properties of mesoscopic conductors
can be understood in terms of noninteracting electrons.
Nevertheless, electron-electron interactions (EEI) are a crucial ingredient
to describe e.g. Luttinger liquids \cite{Kane92},
persistent currents in mesoscopic metallic rings \cite{Berk96}, 
or metal-insulator transition (in 2D systems) \cite{Puda97}.  

In the metallic regime, the conductance of a phase-coherent disordered 1D 
conductor is reduced below its classical Drude value due to EEI \cite{review_ee}.
This Coulomb interaction contribution corresponds to
$\delta G_{EEI} \simeq -(g_F-g_H) e^2/h $ where $g_H \leq g_F\simeq1$.
The screening-dependent Hartree constant $g_H$ is of order unity 
and reaches a minimum (maximum) for long-range (short-range) EEI.
The parallel-spin (exchange) Fock constant $g_F$ is universal
and independent of screening.
Furthermore, $g_F$ and $g_H$ can be subdivided into a diffuson and a 
cooperon contribution due to two electrons travelling along a closed path 
in the same or opposite direction respectively.

A perpendicular magnetic flux of about one flux quantum $h/e$ destroys
the cooperon contribution,
whereas the diffuson contribution is predicted to be {\em insensitive} 
to a perpendicular magnetic flux.
In the presence of a {\it parallel} magnetic field,
spin-up electrons are Zeeman splitted from the spin-down electrons,
which does not affect the (exchange) Fock and parallel-spin 
Hartree contributions.
However, the anti-parallel spin Hartree contribution is expected to be reduced 
when the Zeeman energy $E_Z= g \mu_B B$ exceeds the 
Thouless energy $E_T= \hbar D/L^2$, where $L$ denotes the 
length of the conductor and $D$ the diffusion constant. 
Therefore,  a parallel magnetic field is predicted to enhance $\delta G_{EEI}$
\cite{Kawa81}, which results in a {\em positive} magnetoresistance.
The above prediction also holds in the presence of 
spin-orbit interaction \cite{Kawa81,Nitt97}.

Experimentally, the interaction contribution has been observed in the 
conductance via its temperature-dependence \cite{Uren80,Lin84} or via a 
negative parabolic magnetoresistance in perpendicular fields \cite{Paal83}.
In principle, $\delta G_{EEI}$ could be masked by two other
quantum interference contributions arising from noninteracting electrons:
universal conductance fluctuations (UCF) and 
weak localization (WL) \cite{review_ucf}.
In the above mentioned experiments, a magnetic flux of $h/e$ was used 
to destroy WL and macroscopic conductors were used to suppress UCF 
by ensemble averaging.

A positive magnetoresistance in parallel field was observed in Si:P and 
Si MOSFETs \cite{Rose81}. 
However, the magnitude of $\delta G_{EEI}$ was found to be much larger than 
$e^2/h$.
Recently, it has been shown that a parallel magnetic field drives the 
anomalous conducting phase in Si MOSFETs into the insulating state 
\cite{Puda97}.
This asks for a reinterpretation of the observations reported 
in Ref.~\cite{Rose81}. 
The magnetoresistance of a macroscopic GaAs/AlGaAs 2-dimensional electron 
gas (2DEG) was studied by Lin {\it et al.} \cite{Lin84} in parallel field. 
Although the interaction contribution was present, 
they observed a {\it negative} (temperature-dependent) 
magnetoresistance in parallel field.
This apparent contradiction with the theoretical expectation led these
authors to disregard the Coulomb interaction effect as an explanation.

The experimental status about the behavior of the interaction 
contribution to the conductance, in particular its  
dependence in parallel magnetic fields, is confusing.
Here, we report a study of the interaction contribution in
a phase-coherent disordered 2DEG in which we vary the resistance over
an order of magnitude with an overall gate. 
First, we identify the interaction contribution by applying a 
bias voltage to suppress all phase-coherent contributions.
We unambiguously demonstrate that the magnitude of the  
interaction contribution corresponds to 
$\delta G_{EEI}\simeq-0.3 e^2/h$ independent of resistivity.    
Secondly, we explicitly study its magnitude in parallel magnetic fields
and find a negative magnetoresistance.
Thirdly, fluctuations in the phase-coherent contribution to the 
zero-bias magnetoresistance in perpendicular field are observed 
on a field scale much larger than that expected for UCF,
which might be caused by fluctuations in the interaction contribution.

\begin{figure}[tb]
\centerline{\psfig{figure=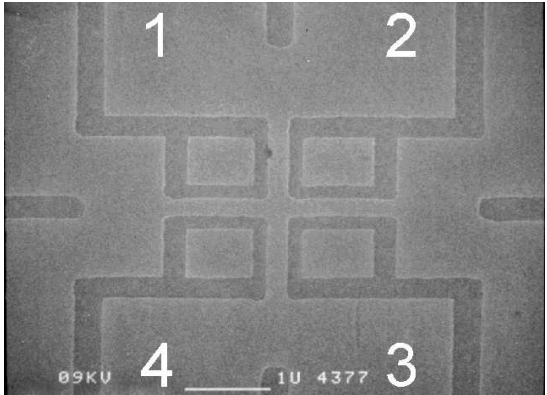,width= 0.4\textwidth}}
\caption[]{Scanning electron micrograph of the 
disordered cross-shaped 2DEG before depositing the gate-electrode.
The darker regions represents insulating trenches in the 2DEG.}
\label{fig1}
\end{figure}

The 2DEG is present in an InAs/AlSb quantum well.
Prior to processing, the top barrier has been removed by wet chemical etching.
The 15nm thick exposed InAs layer hosts the 2DEG with an electron density 
$n_s \simeq 1.5 \cdot 10^{16}$m$^{-2}$ and
an electron mean free path $\ell_e \simeq 0.2\mu$m.
The cross-shaped pattern in the InAs-layer was defined by insulating 
trenches using e-beam lithography and wet chemical etching. 
Note that its length $L\simeq2.1\mu$m and width $W\simeq0.35\mu$m 
are larger than $\ell_e$, 
which implies that transport is diffusive (at zero magnetic field).
After taking scanning electron micrographs (Fig.~\ref{fig1}), 
a 65 nm SiO$_2$ layer (PECVD) and a 40 nm Ti/Au electrode are deposited.
The Ti/Au electrode covers the entire area displayed in Fig.~\ref{fig1}.

\begin{figure}[tb]
\centerline{\psfig{figure=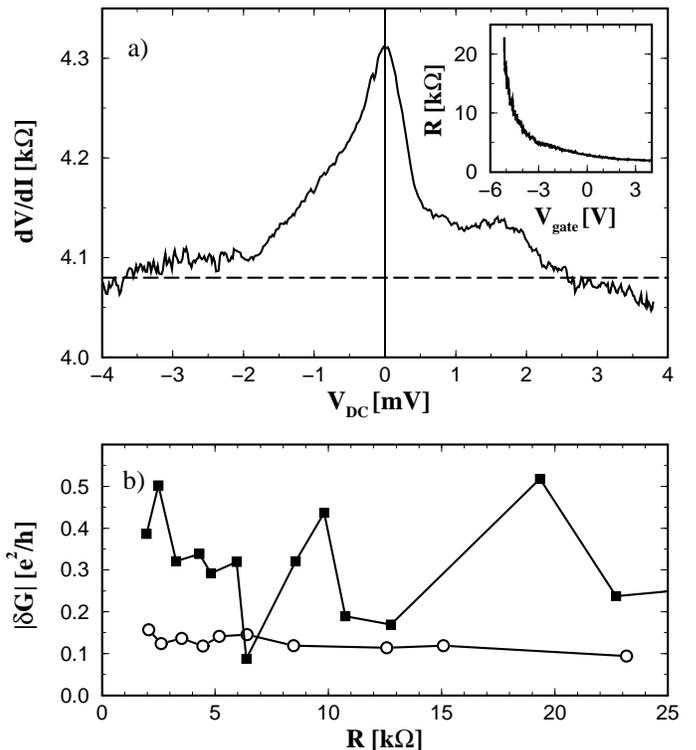,width= 0.5\textwidth}}
\caption[]{Panel a) shows the differential resistance $R_{14,23}$ 
versus applied $dc$ bias voltage measured at $V_{gate}$=-1.5V, 
T=140mK, and a perpendicular magnetic field of 1T.
The dashed line represents the (average) classical resistance 
determined at $\pm$3mV.  
The inset shows the resistance ${\rm R}_{14,23}$ 
at zero bias versus gate voltage.
The filled squares displayed in panel b) represent
the absolute magnitude of the reduction in zero-bias conductance 
compared to the classical conductance  
versus zero-bias resistance.
The open circles denotes the rms magnitude of the (universal)
magnetoconductance fluctuations obtained over a magnetic field range 
between 0.5 to 4.5T.}
\label{fig2}
\end{figure}

We have studied four nominally identical devices at low temperatures.
The differential resistance R$_{14,23}$ is measured
by applying an $ac$ (and $dc$) current between terminals 1 and 4 and measure
the $ac$ voltage between terminal 2 and 3 with a lock-in technique.
The gate voltage is applied with reference to one of the  
terminals connected to the 2DEG. 
The resulting gate-voltage dependence of R$_{14,23}$ is plotted
in the insert of Fig.~\ref{fig2} a). 
The side-contacts were used to monitor $n_s$ versus gate voltage,
which showed that depletion occured at about -5.5V.
For negative gate-voltages only the first 2D-subband of the 2DEG is populated.
The Coulomb interaction range remains in this experiment larger than 
the Fermi wavelength (long-ranged), which results in an maximum magnitude
of $\delta G_{EEI}$ (small $g_H$).

The differential resistance R$_{14,23}$ at 140mK and 1T is
displayed in Fig.~\ref{fig2} a) versus $dc$ bias voltage.
The resistance at zero bias is clearly enhanced compared to at high bias.
In these devices, the magnetoresistance around B=0 
did not reveal a clear signature of WL and the cooperon  interaction
contribution on a scale of a flux quantum $h/e$ through the cross-shaped 2DEG,
which implies that both cooperon contributions are negligible.
Nevertheless, we applied a perpendicular magnetic field of 1T,
much larger than $h/e$ to assure that both WL and the 
cooperon interaction contribution are eliminated. 
Consequently, the only remaining phase-coherent contributions are UCF and 
the diffuson interaction contribution to the conductance.

When the bias voltage (or temperature) is increased the interaction
contribution should be suppressed.
When  $\ell_{T,V}=\sqrt{\hbar D/\mbox{\rm max}(k_BT,eV)}$ is larger than $W$, 
it should show a square root energy dependence 
$\delta G_{EEI}\sim \ell_{T,V}/W$ (1D EEI).
At higher bias voltages, ($V_{DC}\simeq$1mV) $\ell_{T,V}$ becomes smaller 
than $W$ and the interaction contribution is expected to vanish 
logarithmically $\delta G_{EEI}\sim \ln(\ell_{T,V}/\ell_e)$ (2D EEI),
which would eventually only leave UCF around the classical resistance.
We observed, however, that UCF was suppressed at high biases,
which means that the applied bias voltage also destroys phase coherence.
Note that UCF is responsible for the asymmetry in the resistance 
around zero bias voltage. 
We define experimentally the classical resistance as the average of the
resistance at a bias of about $\pm$3mV.
The phase-coherent resistance contribution is thus equal to the difference  
between the zero-bias and classical resistance. 
The resulting contribution to the conductance is plotted in Fig.~\ref{fig2} b)
as a function of resistivity by changing the gate voltage.
This contribution still includes UCF, which can be eliminated by
ensemble averaging.  
Therefore, the average magnitude $\delta G_{EEI} \simeq -0.3 e^2/h$  
of the values plotted in Fig.~\ref{fig2} b) 
unambigously corresponds to the diffuson interaction contribution 
to the conductance \cite{note}.
The observed deviations from this average magnitude corresponds well to the 
contribution due to the magnetoconductance fluctuations (CF). 
The rms magnitude of these CF, plotted in Fig.~\ref{fig2} b), 
is $\delta G_{CF} \simeq 0.13 e^2/h$ independent of the resistance.

\begin{figure}[tb]
\centerline{\psfig{figure=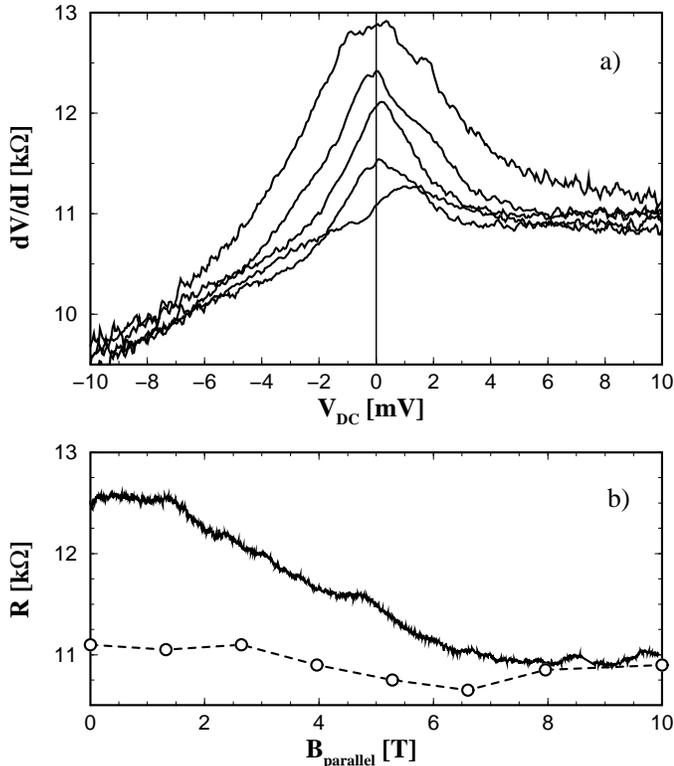,width= 0.5\textwidth}}
\caption[]{Panel a) displays $R_{14,23}$ versus $dc$ bias voltage 
measured at fixed {\em parallel} magnetic fields: 
from top to bottom 0, 2.5, 4, 5.5, and 10T, 
with $V_{gate}$=-4.0V and $T$=1.7K.  
Panel b) shows the zero bias magnetoresistance versus parallel 
magnetic field (solid line). 
The circles represent the classical resistance determined at high-bias (+8mV) 
from R(V)-curves displayed in panel a).}
\label{fig3}
\end{figure}

After having characterized the interaction contribution 
in our phase-coherent conductor, we continue by investigating its 
{\it parallel} magnetic field dependence 
(applied in the direction of the current flow).
The bias-voltage dependence of the differential resistance $R_{14,23}$
at 1.7K is plotted in Fig.\ref{fig3} a) for increasing magnetic fields 
applied to a 13$^\circ$ tilted device.
This temperature of 1.7K implies that $\ell_T\simeq 0.3\mu$m$<W$ (2D EEI),
which could be responsible for the slower bias voltage dependence 
of the interaction contribution compared to that shown in Fig.~\ref{fig2} a).
If the cooperon interaction contribution and WL would have been present 
in our devices, 
they should have been suppressed by the perpendicular component of the applied
magnetic field above 0.2 T.  
Instead of showing the theoretically predicted enhancement for increasing 
magnetic fields, 
this diffuson interaction contribution is suppressed.
In panel b), the zero-bias resistance is plotted versus parallel magnetic
field, which exhibits a {\it negative magnetoresistance}. 
Comparison with the magnetic-field-independent classical resistance reveals 
that the interaction contribution vanishes around a 
magnetic field of 7T.
An estimation for the Zeeman energy at 7T is about  
$E_Z\simeq 1.6$meV ($\gg k_BT\simeq 0.15$meV) using a 
$g$-factor of -4 \cite{Mend94}.
This corresponds to an applied bias voltage where the interaction contribution 
is substantially suppressed. 
The data of Fig.~\ref{fig3} shows that the diffuson interaction contribution
is destroyed by the Zeeman energy, which is in apparent conflict with 
theoretical predictions.
In retrospect, we suggest that the negative magnetoresistance 
in parallel fields observed by Lin {\it et al.} \cite{Lin84} is also caused by 
a suppression of $\delta G_{EEI} \simeq - 0.8e^2/h$ due to Zeeman splitting. 

Apart from the predicted increase at zero bias of the magnitude
of the interaction contribution, a reduction in the interaction contribution
has been predicted when the applied bias voltage becomes equal to the 
Zeeman energy \cite{review_ee}.
For these bias voltages, the anti-parallel spin Hartree contribution
is (partially) restored and, consequently, 
the magnitude of $\delta G_{EEI}$ is reduced.
This prediction was actually the original motivation to study the 
bias-voltage dependence in parallel magnetic fields.
However, these Zeeman Coulomb interaction contributions around bias 
voltages $eV_{DC}=\pm E_Z$ are not present in our devices 
as can be verified in Fig.~\ref{fig3} a).

\begin{figure}[thb]
\centerline{\psfig{figure=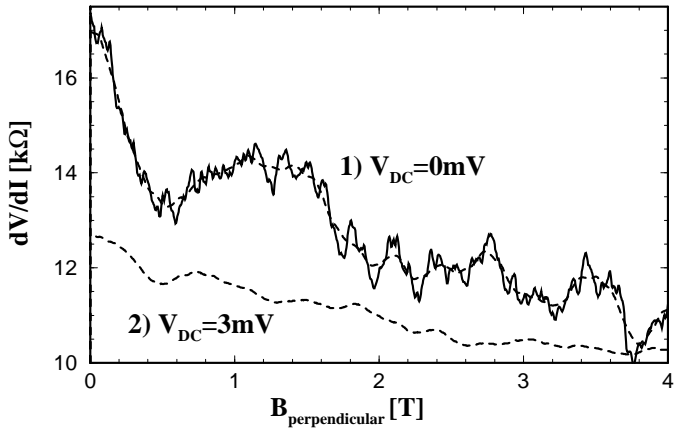,width= 0.5 \textwidth}}
\caption[]{The magnetoresistance $R_{14,23}$ in 
{\em perpendicular} magnetic field at $V_{gate}$=-4.7V and T=180mK.
The solid line 1) displays the zero-bias resistance which exhibits 
conductance fluctuations.
The dashed lines 1)  and 2) display the resistance averaged over 
$\Delta$B=0.2T for a bias voltage of 0 and 3mV respectively,
which difference should indicate the Coulomb interaction contribution.}
\label{fig4}
\end{figure}

The striking parallel magnetic field dependence of the interaction
contribution motivated us to thoroughly study its sensitivity to a 
perpendicular magnetic field. 
The onset of Shubnikov-de Haas oscillations in high magnetic 
fields limits us to magnetic fields below about 4 T.
The reproducible magnetoresistance traces measured at 180mK with a bias 
voltage of 0 and 3mV are plotted in Fig.~\ref{fig4}. 
We checked that at a bias of 3mV the interaction contribution 
was almost completely suppressed (see Fig.~\ref{fig2} a)).
We believe that the (relatively small) conductance fluctuations with a 
magnetic field sensitivity of about $B_c\simeq$0.03 T 
($\simeq 10 h/e$ through total cross-shaped 2DEG) are related to UCF,
which can be suppressed by averaging over 0.2T.
The solid line represents the magnetoresistance  
measured at zero bias without averaging.
The difference between the two dashed lines should directly
reflect the interaction contribution.
Its magnitude averaged between 0 and 2.5 T is 
$\delta G_{EEI} \simeq -0.3 e^2/h$ independent of the resistivity,
which is similar to the value found in Fig.~\ref{fig1} b).

The gradual reduction of the interaction contribution by a perpendicular 
magnetic field is related to the observed suppression by the Zeeman energy.
In addition, relatively pronounced fluctuations are observed
with a large magnetic field scale of about 1 T,
which are absent in the classical high-bias resistance.
This behavior is observed for all applied gate voltages.
Note that the rms magnitude of the conductance fluctuations plotted in 
Fig.~\ref{fig1} b) is mainly determined by these pronounced fluctuations.
At first instance, one would attribute these fluctuations to UCF,
which rms magnitude and magnetic field sensitivity are 
expected to be unaffected by EEI \cite{Lee87}.
However, these pronounced fluctuations occur at a
magnetic field scale where the cyclotron radius $\ell_c$ becomes 
comparable to the elastic mean free path $\ell_e$. 
Although the typical field scale of UCF can be enhanced by a factor of 
about $\sqrt{1+(\ell_e/\ell_c)^2}$ \cite{Xion92}, 
the observed field scale of about 1 T is much larger and cannot  
be reconciled with the present UCF theory.
As an alternative explanation we suggest that this might be due to 
fluctuations in the interaction contribution. 

In conclusion, we have studied the Coulomb interaction contribution
to the conductance of a cross-shaped phase-coherent disordered 2DEG.
The interaction contribution was found to be suppressed by a parallel 
magnetic field, which results in a negative, instead of the predicted 
positive, magnetoresistance.
In perpendicular magnetic fields, reproducible fluctuations in the 
magnetoresistance were observed, which seems to be inconsistent with UCF and
might be related to fluctuations in the interaction contribution.

We have benefitted from discussions with J.P. Heida and A.F. Morpurgo
about the persistant presence of a zero-bias anomaly in the resistance.
This work is part of the research program of the stichting voor
Fundamenteel Onderzoek der Materie (FOM), which is financially supported
by the Nederlandse organisatie voor Wetenschappelijk Onderzoek (NWO).
B.J. van Wees acknowledges support from the Royal Dutch Academy of
Sciences (KNAW).

\bibliographystyle{prsty} 

\begin{thebibliography}{10}
\bibitem{note}
A small decrease of $\delta G_{EEI}$ could be expected 
when the resistivity increases (EEI remains long ranged).
\end{thebibliography}

\end{document}